\documentclass{itor}

\usepackage{natbib}%
\usepackage[figuresright]{rotating}

\usepackage{url}
\usepackage{algorithmic}
\usepackage{algorithm}

\usepackage{amsmath,amsthm}

\theoremstyle{definition}

\theoremstyle{remark}

\usepackage{hyperref}
\usepackage{caption}
\usepackage{multirow}
\usepackage{pgfplots}
\usepackage{tikz}

\begin{document}

\title{Cascading-Tree Algorithm for the 0-1 Knapsack Problem\\ 
\normalsize{In Memory of Heiner Müller-Merbach, a Former President of IFORS}
}

\author[Running Author]{Mahdi Moeini\affmark{a,$\ast$}, Daniel Schermer\affmark{c} and Oliver Wendt\affmark{c}}

\affil{\affmark{a}ENSIIE, 1 Place de la Résistance, 91000 Évry-Courcouronnes, France.}
\affil{\affmark{c}Chair of Business Information Systems and Operations Research (BISOR), 
RPTU Kaiserslautern-Landau, 67663 Kaiserslautern, Germany.}
\email{mahdi.moeini@ensiie.fr [Mahdi Moeini]; daniel.schermer@wiwi.uni-kl.de [Daniel Schermer]; \\ wendt@wiwi.uni-kl.de [Oliver Wendt]}

\thanks{\affmark{$\ast$}Author to whom all correspondence should be addressed (e-mail: mahdi.moeini@ensiie.fr).}

\historydate{}

\begin{abstract}
In operations research, the Knapsack Problem (KP) is one of the classical optimization problems that has been widely studied. The KP has several variants and, in this paper, we address the binary KP, where for a given knapsack (with limited capacity) as well as a number of items, each of them has its own weight (volume or cost) and value, the objective consists in finding a selection of items such that the total value of the selected items is maximized and the capacity limit of the knapsack is respected. In this paper, in memorial of Prof.\ Dr.\ Heiner Müller-Merbach, a former president of IFORS, we address the binary KP and revisit a classical algorithm, named \textit{cascading-tree branch-and-bound algorithm}, that was originally introduced by him in 1978. However, the algorithm is surprisingly absent from the scientific literature because the paper was published in a German journal. We carried out computational experiments in order to compare the algorithm versus some classic methods. The numerical results show the effectiveness of the interesting idea used in the cascading-tree algorithm. 
\end{abstract}


\maketitle

%
\section{Introduction}
\label{Section:Introduction}
In operations research (OR), it is not surprising to deal often with NP-hard problems. A classical example of such problems is the well-known \textit{Knapsack Problem} and its variants, e.g., the \textit{0-1 Knapsack Problem} (KP) 
\citep{Ghosh.2001,BookSpringer.2004,Book.MartelloToth.1990,Pisinger.1995,Book.Zimmermann.2008}. In the KP, we assume that a knapsack and a set of items are given, where each item has its own weight (volume or cost) as well as value. Taking into consideration the capacity limit of the knapsack, we want to find a selection of items such that the total value of the selected items is maximized.

Assume that $n$ items (objects) are given such that item $j$ has the weight $w_j$ and value $v_j$, where $j \in\{1,\dots,n\}$.
We suppose that the items are sorted in the decreasing order of $v_j / w_j$, i.e., 
\[ %
\frac{v_1}{w_1} \geq \dots \geq \frac{v_n}{w_n}. %
\] %
Let $C$ denote the capacity of the knapsack. Without loss of generality, we assume that all input data are positive numbers.
Define the binary variable $x_j$ to be $1$ if and only if item $j$ is selected to be in the knapsack, and $0$ otherwise. Then, the 0-1 Knapsack Problem is mathematically formulated as follows:

\begin{equation}
\max \left\{
\sum\limits_{j=1}^{n} v_{j}x_{j} \quad | \quad \sum\limits_{j=1}^{n} w_{j}x_{j}\leq C, x_j \in \{0,1\} : j \in \{1,\dots,n\}
 \right\}. \label{eq4:KP-model}
\end{equation}%

The KP has many applications in industrial and academic problems, e.g., project selection, cargo loading, cutting stock, and budget control, where either the problem is formulated as a KP or it is used as a subproblem through an algorithm \citep{BookSpringer.2004,Book.MartelloToth.1990}. 

The KP has several variants, e.g., the \textit{bounded knapsack problem}, \textit{subset-sum problem}, \textit{0-1 multiple knapsack problem}, among others. Each variant is based on some specific assumptions and/or generalization of the KP. The KP and its variants have been widely studied in operations research. 
For a detailed literature review, the interested reader can refer to \citep{BookSpringer.2004,Dudzinski.1987,Book.MartelloToth.1990,Salkin.1975}, and references therein. In the following, we review briefly some of the well-known approaches that have been introduced for solving the KP.

In addressing the KP, several exact or heuristic algorithms have been used in the scientific literature \citep{Parada.2016}. Indeed, since the KP belongs to the group of NP-hard problems in the \textit{weak} sense, the KP can actually be solved in pseudopolynomial time \citep{BookSpringer.2004,Martello.1999}. Hence, as exact solution approach, Branch-and-Bound and Dynamic Programming, with exponential and pseudopolynomial complexity, respectively, are well suited for solving the KP. At the current state of the art, a considerably large-scale instances of the KP can be solved exactly in a reasonable computation time \citep{BookSpringer.2004,Book.MartelloToth.1990}. The sophisticated exact approaches use tight bounds, valid inequalities, and variable reduction techniques with the purpose of enhancing the performance of the algorithm (see, e.g., \citep{BookSpringer.2004,Book.MartelloToth.1990,Martello.2000}). 

In this context, the branch-and-bound paradigm was always considered as a successful approach in solving the KP, e.g., \citep{Greenberg.1970,Kolesar.1967,Martello.2000,HMM.CascadingTree.1978}. Even though most of these approaches are quite well-known in the literature of KP \citep{BookSpringer.2004,Book.MartelloToth.1990}, the \textit{cascading-tree algorithm} (CTA) is absent from the literature, possibly, because of the fact that the corresponding paper was originally published in German \citep{HMM.CascadingTree.1978}. However, the CTA uses a smart branching strategy that has the potential of increasing its performance. Hence, we dedicate the current memorial paper to two objectives: We revisit the CTA and compare it with two classical branch-and-bound algorithms in solving the 0-1 knapsack problem. We selected the branch-and-bound algorithms introduced by Kolesar \citep{Kolesar.1967} and Greenberg \& Hegerich \citep{Greenberg.1970} as they fit well to the context of this paper. Even though there are currently quite more efficient and well-elaborated branch-and-bound algorithms for solving the KP, this paper does not aim at introducing a new innovative and more efficient algorithm.
In order to show the impact of the cascading branching strategy, we have conducted computational experiments on randomly generated instances \citep{Martello.1999} and report the numerical results. 
In addition, as a memorial paper, we keep alive the memory of Prof.\ Dr.\ Heiner Müller-Merbach, a former president of IFORS, by presenting his contributions to the OR community. 

The remainder of this paper is organized as follows.
We begin in Section \ref{Section:HMM} by providing an overview on the contributions of Prof.\ Dr.\ Heiner Müller-Merbach. 
Afterwards, in Section \ref{Section:algorithms}, we explain briefly the selected branch-and-bound algorithms for solving the 0-1 knapsack problem. 
Section \ref{Section:computationalExperiments} is devoted to the numerical studies of our investigations.
Finally, concluding remarks on this work and future research implications are provided in Section \ref{Section:conclusion}.

%
\section{Heiner Müller-Merbach: Short Biography and Scientific Contributions}
\label{Section:HMM}
%
In this section, we provide a concise biography of \textit{Prof.\ Dr.\ Heiner Müller-Merbach}.

Prof.\ Dr.\ Heiner Müller-Merbach was born on June 28, 1936, at Hamburg in Germany. He spent his childhood there and completed his studies until he graduated from high-school in 1955.
Between 1955 and 1960, Heiner Müller-Merbach studied Business Administration and Management Engineering (Wirtschaftsingenieurwesen) at the Technische Universität Darmstadt. In 1962, he obtained a PhD and in 1967 his Habilitation (PhD II) from the Technische Universität Darmstadt.
Between 1961 and 1967, he was lecturer and researcher at the ``Institute für Praktische Mathematik" at the Technische Universität Darmstadt. Within this period, in 1963, he had a postdoctoral program at the Operation Research Centre of the University of California Berkeley, where he worked with George B. Dantzig, the founder of the well-known simplex algorithm.
Between 1967 and 1971, he taught Business Administration at the University of Mainz and then, he received the full professor position in Operations Research at the Technische Universität Darmstadt. He had this position until 1983. Afterwards, he moved to the \textit{Technische Universität Kaiserslautern} as the professor in Business Information and Operation Research. In 2004, he became an emeritus professor. 

The research interest theme of Heiner Müller-Merbach cover mainly linear programming, combinatorial optimization, branch-and-bound algorithms, and heuristics. Later, in the 90s, he became interested in organizational intelligence and knowledge management as well as in leadership and philosophy.

Heiner Müller-Merbach has played an important role in the \textit{International Federation of Operational Research Societies} (IFORS). First, he was vice-president from 1974 to 1976 and then, he became president from 1983 to 1985. He also participated to the organization of the IFORS international Conference and was a member of the Organizing Committee of the 9th International IFRS Conference that took place in Hamburg.
Moreover, he had some honorary positions in several other organizations such as the \textit{Institute of Management Sciences}.

Heiner Müller-Merbach published a large number of books and papers, e.g.,
\citep{HMM.1961,HMM.1966a,HMM.1966b,HMM.1966c,HMM.LNE.1970,Book.HMM.1970,HMM.EOR.1970,HMM.1973,Book.HMM.1973,HMM.1975,HMM.ZOR.1976,HMM.1978,HMM.CascadingTree.1978,HMM.1981}. Overall, he wrote 14 books. A non comprehensive list is  as follows:
\begin{itemize}
\item \textit{Operations Research}. Munich, (basic OR/MS textbook in German, 565 pages) \citep{Book.HMM.1973}. 
\item \textit{Optimale Reihenfolgen} (In English: Optimal Sequences). Springer. Berlin 1970 (monograph on combinatorial sequencing problems, 225 pages) \citep{HMM.EOR.1970}. 
\item \textit{Operations-Research-Fibel für Manager} (In English: OR/MS primer for managers). Moderne Industrie. Munich (108 pages) \citep{Book.HMM.1970}. 
\item \textit{On Round-Off Errors in Linear Programming}. Springer. Berlin 1970 (monograph on round-off error growth in LP codes) \citep{HMM.LNE.1970}. 
\end{itemize}

He also published about 400 articles in different journals, conference proceedings, and collective books. Here is a quite short list of selected papers:
\begin{itemize}
\item \textit{An Improved Starting Algorithm for the Ford-Fulkerson Approach to the Transportation Problem}, published in \textit{Management Science} \citep{HMM.1966a}.
\item \textit{Upper-Bounding-Technique, Generalized Upper-Bounding-Technique and Direct Decomposition in Linear Programming: A Survey on Their General Principles Including a Report about Numerical Experience}, In: \textit{Decomposition of Large-Scale Problems} (Ed. David M. Himmelblau), North-Holland, Amsterdam 1973, pp. 167-180. \citep{HMM.1972}
\item \textit{Improved upper bound for the zero-one knapsack problem. A note
on the paper by Martello and Toth}, European Journal of Operational Research, 3:212-213,1978 \citep{HMM.1978}
\end{itemize}

Heiner Müller-Merbach was in the editorial advisory boards of 12 journals, e.g., the \textit{European Journal of Operational Research} (1977 – 2000) and the International Journal of General Systems (1974-1994). He was the Editor-in-Chief of \textit{Technologie \& Management} from 1985 to 1997.
Finally, between 2006 and 2009, he was the scientific director of \textit{Foundation for International Business Administration Accreditation} (FIBAA) in Germany. 

Professor Heiner Müller-Merbach passed away at the age of 78 on May 30, 2015, in Darmstadt (Germany).

%
\section{Branch-and-Bound Algorithms for the 0-1 Knapsack Problem}
\label{Section:algorithms}
%
In this section, we present three branch-and-bound algorithms for solving the 0-1 knapsack problem. As an implicit enumeration approach, any branch-and-bound algorithm is based on two fundamental principals: \textit{branching} and \textit{bounding}. The branching part is used to divide the solution space into smaller ones, and the objective bounding operation consists in finding lower and upper bounds on the optimal solution value of the problem (for more details, refer to \citep{BookSpringer.2004}). Two different branch-and-bound algorithms might basically differ in way that they implement the \textit{branching} and \textit{bounding} steps.

In this paper, we will focus on solving the classical 0-1 Knapsack Problem, and use the algorithms of Kolesar, Greenberg \& Hegerich, and the cascading-tree \citep{Greenberg.1970,Kolesar.1967,HMM.CascadingTree.1978}. The algorithm of Kolesar is probably the first branch-and-bound algorithm introduced for solving the 0-1 knapsack problem. Even though, quite efficient algorithms can solve very large-scale KP instances \citep{BookSpringer.2004,Book.MartelloToth.1990}, these basic algorithms fit well to the objectives of the the current memorial paper in which we want to investigate the impact of the cascading branching strategy. 

In these basic algorithms, any feasible integer solution defines a lower bound, and an upper bound is obtained through relaxation of the binary restrictions of the variables to real-valued variables between $0$ and $1$. Despite Kolesar, Greenberg \& Hegerich algorithms, in the cascading-tree algorithm, Heiner Müller-Merbach suggests a lower bound through a heuristic method.
The main difference remains indeed in different branching strategies (refer to Sections \ref{Section:algorithm-ClassicAlgos} and \ref{Section:algorithm-CascadingTree}).

\subsection{Classic Algorithms for the 0-1 Knapsack Problem}
\label{Section:algorithm-ClassicAlgos}

In 1967, Kolesar introduced the first branch and bound approach which is an exact solution of the Knapsack Problem \citep{Kolesar.1967}. Once the solution of the relaxed problem is obtained, the algorithm starts branching from highest variable index, i.e., $x_1$, by fixing it to $0$ or $1$. The subsequent branching variables are selected in decreasing order of the variable indices. Each subproblem is evaluated by relaxing the binary conditions. While looking for a subproblem for further branching, the open subproblem with the highest upper bound is selected (ties broken arbitrarily). The lower bound is updated as soon as any new integer feasible solution is found. The algorithm stops only if there is no way to find a better solution. 

In 1970, Greenberg and Hegerich, proposed a different approach for solving the Knapsack Problem \citep{Greenberg.1970}. Their algorithm has its main difference with the Kolesar's algorithm in the selection of branching variable. More precisely, the algorithm of Greenberg \& Hegerich selects the fractional variable for branching on.

\subsection{Algorithm of Cascading-Tree}
\label{Section:algorithm-CascadingTree}

According to the article by Müller Merbach \citep{HMM.CascadingTree.1978}, the cascading-tree algorithm is composed of four basic principles. 
The first one is the selection of a node, which is typically an open node, i.e., not yet explored, with highest upper bound. The second principle is how to branch on the selected node. The cascading-tree algorithm suggests a simultaneous branching approach from which the term ``cascading" stems. More precisely, once a node is created, the algorithm assigns to it a heuristic integer solution, which is explained in the following, and uses this integer solution for branching. Assume that a node with an integer solution vector $x:=(x_1, \dots, x_i, \dots, x_n)$ is given on which the cascading branching should be performed. We use all variables $x_i$ that are equal to $1$ for branching (under the condition of respecting the capacity constraint of the knapsack). For this purpose, assume that the vector $x$ contains a sequence (not necessarily consecutive) of variables $x_{i_1}, x_{i_2}, \dots, x_{i_r}$ that are equal to $1$. In addition, we assume that none of these variables have already been set to $0$ or $1$ (the following description can easily be extended to cover the case with already fixed variables). 
\begin{itemize}
\item Starting from $x_{i_1}$, the first branch sets $x_{i_1} \leftarrow 0$, 
\item the second branch sets $x_{i_1} \leftarrow 1$ and $x_{i_2} \leftarrow 0$, 
\item the third branch sets $x_{i_1}=x_{i_2} \leftarrow 1$ and $x_{i_3} \leftarrow 0$,
\item the $r$-th branch sets  $x_{i_1}= \ldots =x_{r-1} \leftarrow 1$ and $x_r \leftarrow 0$.
\end{itemize}

In addition, after branching, if the weight of an item is larger than the remaining capacity, the variable corresponding to the item can be set to zero branch for the corresponding subproblem.
This style of branching cascades on setting variables and incorporates two natural intuitions.
On the one hand, higher branches (i.e., closer to $r$) might yield better relaxations, as the values with the largest value per weight ratio are not fixed to zero.
As a result, these branches might be more valuable to explore.
On the other hand, these higher branches also contain a smaller subset of solutions, as a larger amount of variables are fixed.
Therefore, they might be searched more effectively.

\begin{algorithm}[H]
\begin{algorithmic}[1]
\STATE Procedure for solving the knapsack problem $P^0$;
\STATE List of open subproblems $L \leftarrow P^0$;
\STATE Upper bound $\leftarrow$ $-\infty$, lower bound $\leftarrow$ $0$;
\WHILE{($\neg$ Stopping Criteria)}
   \STATE Select a problem $P \in L$ with the highest upper bound;
   \STATE Solve the relaxed problem of $P$ to find solution vector $x^P$ and its corresponding objective value $f(x^P)$;
   \IF{$x^P$ is integer}
            	\STATE $\overline{x^P} \leftarrow x^P$;
            	\IF{$f(x^P)$ $\geq$ lower bound}
					\STATE lower bound $\leftarrow f(x^P)$;
				\ENDIF   
			\ELSE 
				\STATE Compute heuristic integer solution $\overline{x^P}$ and bound $f(\overline{x^P})$; 
				\IF{$f(\overline{x^P})$ $\geq$ lower bound}
					\STATE lower bound $\leftarrow f(\overline{x^P})$;
				\ENDIF 
				\STATE Execute cascade branching and create subproblems $P^1, P^2, \dots, P^l$;
				\STATE Add all feasible $P^i$ to $L$ ($1\leq i \leq l$);
			\ENDIF
        \ENDWHILE
        \STATE \textbf{return} solution;
\end{algorithmic}
\caption{Cascading-Tree Algorithm}  %
\label{algo:cascading-tree} %
\end{algorithm}

The third principle corresponds to bounding, i.e., each newly formed solution must be characterized by two limits: an upper bound and a lower bound. In a similar way to the algorithms of Kolesar as well as Greenberg \& Hegerich, upper bound is the Dantzig bound \citep{Dantzig.1957,BookSpringer.2004,Book.MartelloToth.1990}, i.e., it is determined by solving the relaxed problem. However, to compute lower bounds, the cascading-tree algorithm uses a different approach than using simply any integer feasible solution. In fact, the cascading-tree algorithm allows heuristics for generating integer solutions.
As one possible heuristic, assume that $x:=(x_1, \dots, x_i, \dots, x_n)$ is a current node, which is selected for branching. Without loss of generality, we can assume that $x_1 > 0$. Assume that $x_i$ is the variable with fractional value.  
Then, we proceed as follows: in the heuristic integer solution, denoted by $\overline{x}$, we set $\overline{x}_j \leftarrow x_j$ for all $1 \leq j \leq i-1$, and $\overline{x}_i \leftarrow 0$. Then, starting from $j \leftarrow i+1$, up to $j \leq n$, we set $\overline{x}_j \leftarrow 1$, if and only if the capacity of the knapsack permits. In place of this heuristic, any other one might be used.

Any heuristic integer solution, that is generated in this way, is useful in two aspects: firstly, determining a lower bound is used in termination criteria of the algorithm; secondly, the heuristic integer solution is also used through the cascading branching procedure. 
 
The fourth principle is the termination criteria of the algorithm, where, similar to any other branch-and-bound algorithm, the cascading-tree algorithm stops as soon as there is an evidence on absence of any better integer solution (or feasibility) of the problem. Algorithm 1 summarizes the steps of the cascading-tree algorithm.

\subsection{An illustrative Example}
\label{Section:Example}
In order to illustrate the cascading-tree algorithm, we highlight one of the examples introduced in \cite{HMM.1978}.
Consider Table \ref{tab:example} in which a KP with five items $j \in \{1,\dots,5\}$ with their respective values $v_j$ and weights $w_j$ are shown.

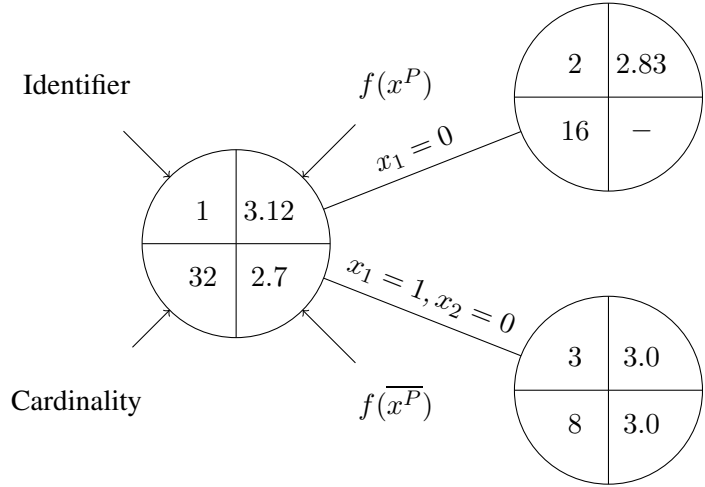
\begin{figure}[]
    \centering
    \begin{tikzpicture}
        \node[circle,draw, minimum width=2.5cm, outer sep=0pt] (a) {};
        \foreach \i in {0,90,180,270}
            \draw (a.center)--(a.\i);
        
        \foreach \i/\j in {45/{3.12},135/{1},225/{32},315/{2.7}}
            \path (a.center) -- node {$\j$} (a.\i); 

        \node[circle,draw, minimum width=2.5cm, outer sep=0pt, above right of=a, node distance=7cm,yshift=-3cm] (b) {};
        \foreach \i in {0,90,180,270}
            \draw (b.center)--(b.\i);
    
        \foreach \i/\j in {45/{2.83},135/{2},225/{16},315/{-}}
            \path (b.center) -- node {$\j$} (b.\i); 

        \node[circle,draw, minimum width=2.5cm, outer sep=0pt, below right of=a, node distance=7cm,yshift=3cm] (c) {};

        \foreach \i in {0,90,180,270}
            \draw (c.center)--(c.\i);

        \foreach \i/\j in {45/{3.0},135/{3},225/{8},315/{3.0}}
            \path (c.center) -- node {$\j$} (c.\i); 
        
        \draw (a)--(b) node[midway,sloped,above] {$x_1=0$};
        \draw (a)--(c) node[midway,sloped,above] {$x_1=1,x_2=0$};

        \node[circle, minimum width=1.5cm, above left of=a, node distance = 3cm] (l1) {Identifier};
        \draw[->] (l1)--(a.135);
        \node[circle, minimum width=1.5cm, above right of=a, node distance = 3cm] (l2) {$f(x^P)$};
        \draw[->] (l2)--(a.45);
        \node[circle,minimum width=1.5cm, below left of=a, node distance = 3cm] (l3) {Cardinality};
        \draw[->] (l3)--(a.225);
        \node[circle,minimum width=1.5cm, below right of=a, node distance = 3cm] (l4) {$f(\overline{x^P})$};
        \draw[->] (l4)--(a.315);
        
        \end{tikzpicture}
    \caption{The application of Algorithm \ref{algo:cascading-tree} on the Knapsack problem portrayed in Table \ref{tab:example}.}
    \label{fig:example}
\end{figure}

Figure \ref{fig:example} illustrates how the cascading-tree algorithm can be used to solve this problem.
The root node, labelled with the identifier 1, contains the set of all possible solutions and, as there are 5 items in this problem, the cardinality is 32.
Following Algorithm \ref{algo:cascading-tree}, we first compute a solution to the relaxed problem at the root node, which yields $x_1=x_2=1$, $x_3 = 0.6$, and $x_4=x_5=0$ with an objective value of $f(x^P)=3.12$.
Moreover, (following the procedure described in Section \ref{Section:algorithm-CascadingTree}) the heuristic solution is $x_1=x_2=1$ and $x_3=x_4=x_5=0$ with an objective value of $f(\overline{x^P})=2.7$.
As a result, the subset of solutions where $x_1=x_2=1$ (there are a total of 8 solutions) does no longer have to be evaluated and cascade-branching is executed, i.e., the branches are created and their respective relaxed solution value is calculated.
In the first branch, labelled with the identifier 2, $x_1 \leftarrow 0$ is fixed and the resulting node contains 16 potential solutions with a relaxed solution value of $f(\overline{x^P})=2.83$.
In the second branch, labelled with the identifier 3, $x_1 \leftarrow 1$ and $x_2 \leftarrow 0$ are fixed which leaves 8 potential solutions with a relaxed solution value of $f(\overline{x^P})=3.0$.
As node 3 has the highest remaining upper bound, it is inspected next.
Notably, the (relaxed) solution of $x_1=x_3=x_4=1$, $x_2=x_5=0$ is already an integer solution.
Therefore, $f(\overline{x^P})=f({x^P})=3.0$ and no further branching is necessary at this node.
By inspection, there is no need to branch on node 2 as its upper bound is less than the objective value of the incumbent solution.
As a result, node 2 can be closed and node 3 contains the optimal solution to the problem.

Noteworthy, solving the same problem using the methods proposed by \cite{Kolesar.1967} and \cite{Greenberg.1970} requires to investigate 11 and 13 nodes, respectively (refer to \cite{HMM.1978}).

\begin{table}
\caption{A sample instance of the knapsack problem with five items $j = 1, \ldots, 5$ as well as their respective value $v_j$ and weight $w_j$ (\cite{HMM.1978}). The items have been sorted according to their value per weight and it is assumed that the capacity is $C=20$.\label{tab:example}}
\begin{tabular*}{\hsize}{@{}@{\extracolsep{\fill}}llllll@{}}
\hline
    $j$     & 1     & 2    & 3    & 4     & 5      \\
    $v_j$    & 1.5   & 1.2  & 0.7  & 0.8   & 0.9    \\
    $w_j$    & 9     & 8    & 5    & 6     & 7      \\
    $v_j/w_j$ & 0.167 & 0.15 & 0.14 & 0.133 & 0.129 \\
\hline
\end{tabular*}
\end{table}

%
\section{Computational Experiments and their Numerical Results}
\label{Section:computationalExperiments}
%

In absence of numerical analysis in \citep{HMM.CascadingTree.1978}, we implemented the three algorithms presented in Section \ref{Section:algorithms} and carried out computational experiments on randomly generated instances, and report the results in this section.

\subsection{Test Settings}
\label{Section:Instances}

Following \citep{Martello.1999}, to evaluate the performance of the presented algorithms, we generated random instances of different sizes. We used three procedures to generate values for $w_j$ and $v_j$, where $R=100$:
\begin{itemize}
\item \textit{Uncorrelated}: for $j \in \{1, \dots, n\}$, $w_j$ and $v_j$ are random numbers uniformly drawn from $[1, R]$.
\item \textit{Weakly correlated}: $w_j$ and $v_j$ values are uniformly random distributed in $[1, R]$ and $[w_j - \frac{R}{10}, w_j + \frac{R}{10}]$, respectively, such that $v_j \geq 1$ and $j \in \{1, \dots, n\}$.
\item \textit{Strongly correlated}: we draw $w_j$ uniformly from $[1, R]$, and set $v_j = w_j+  \frac{R}{10}$, where $j \in \{1, \dots, n\}$.
\end{itemize}
In addition, depending on the $w_j$ values in a given instance, we set $C := \frac{1}{2} \sum\limits_{j=1}^{n} w_j$. 
We generated $5$ instances per each combination of type and size, where $n \in \{10, 20, \dots, 50\}$. 

We implemented the three presented algorithms in Python and tested them, using the $75$ randomly-generated instances, on a Laptop with Intel core i7 CPU and 8 GB RAM. We set a time limit of 3600 seconds on each run. 

\subsection{Numerical Results}
\label{Section:NumericalResults}

We chose to compare them on two criteria: the execution time and the number of nodes computed. Moreover, we have decided to stop the algorithms if the resolution time exceed one hour. 

In solving the instances, we encountered no storage (memory) problems. However, we noticed that the Kolesar and Greenberg \& Hegerich algorithm could not solve some instances in less than one hour, which is the time limit imposed on each run. Indeed, the Kolesar's and Greenberg \& Hegerich's algorithms failed in solving 4 and 5 instances, respectively. We have therefore decided not to take these five instances into account in the calculation of the average node number and computation time.

The results are reported in Tables \ref{tab:Overall}-\ref{tab:averageNodes}. More precisely, Table \ref{tab:Overall} gives an overview on the average number of nodes and the average computation time that each algorithm requires to find the optimum, where the average is taken over all instances.

Tables \ref{tab:averageTime} and \ref{tab:averageNodes} provide more detailed results, where we further specify our findings by comparing two criteria concerning the type and size of the instances. In these tables, each algorithm can be compared to the others line by line. 

As the main observation, according to the numerical results, we see that the cascading-tree algorithm is more efficient that the two other ones. In fact, this shows the importance of the cascading branching strategy because it is the sole difference between three algorithms.

%
\begin{table}
\caption{Overall average results for the 70 instances.\label{tab:Overall}}
\begin{tabular*}{\hsize}{@{}@{\extracolsep{\fill}}llll@{}}
\hline
Algorithm:  & \tch{1}{l}{Kolesar}  & \tch{1}{l}{Greenberg \& Hegerich}  & \tch{1}{l}{Cascading Tree}    \\
\hline 
        Average node number                   
        & 478                                                   
        & 512                            
        & 69              \\
        Average run time (s.)                   
        & 4.078                                                    
        & 22.344                          
        & 0.026            \\
\hline
\end{tabular*}
\end{table}
%

%
\begin{table}
\caption{Comparison on average computation (in seconds) time of each algorithm.\label{tab:averageTime}}
\begin{tabular*}{\hsize}{@{}@{\extracolsep{\fill}}llll@{}}
\hline
Instances  & \tch{1}{l}{Kolesar}  & \tch{1}{l}{Greenberg \& Hegerich}  & \tch{1}{l}{Cascading Tree}    \\
\hline 
        Instances size 10  &   0.025  &  0.032	 &  0.013             \\
        Instances size 20  &   1.175  &  1.422	 &  0.026               \\
        Instances size 30  &   1.346  &  5.250	 &  0.033              \\
        Instances size 40  &   19.684  &  76.138 &  0.025             \\
        Instances size 50  &   0.914  &  37.358	 &  0.034              \\ 
\hline
        Uncorrelated instances   &  0.038  &  0.052	 &  0.021             \\
        Weakly correlated instances  &  0.112  &  0.245	 &  0.022               \\
        Strongly correlated instances  &  12.804  &  77.834	 &  0.037                \\ 
\hline
        \textbf{Average all instances}  &  \textbf{4.078}  &  \textbf{22.344}	 &  \textbf{0.026}                \\
\hline
\end{tabular*}
\end{table}
%

%
\begin{table}
\caption{Comparison on average node number needed by each algorithm.\label{tab:averageNodes}}
\begin{tabular*}{\hsize}{@{}@{\extracolsep{\fill}}llll@{}}
\hline
Instances  & \tch{1}{l}{Kolesar}  & \tch{1}{l}{Greenberg \& Hegerich}  & \tch{1}{l}{Cascading Tree}    \\
\hline 
        Instances size 10  &  113  &  	66	  &  30              \\
        Instances size 20  &  391  &  	258  &  	73                \\
        Instances size 30  &  533  &  	474  &  	91               \\
        Instances size 40  &  869  &  	949  &  	93               \\
        Instances size 50  &  567  &  	893  &  	66               \\ 
\hline
        Uncorrelated instances  &    112	  &  72	  &  32              \\
        Weakly correlated instances  &   269  &  	203	  &  48               \\
        Strongly correlated instances  &    1087	  &  1447	  &  141              \\ 
\hline
        \textbf{Average all instances}  &    \textbf{478}  &  	\textbf{512}	  &  \textbf{69}              \\
\hline
\end{tabular*}
\end{table}
%

%
\section{Conclusion}
\label{Section:conclusion}
%
As a memorial paper, we dedicated it to Professor Heiner Müller-Merbach, who served OR community in IFORS and in academia. In addition, we revisited one of his contributions, i.e., \textit{cascading-tree} approach, which is surprisingly absent from the English scientific literature, consists in a special branching strategy for branch-and-bound algorithms. Using randomly generated instances, we investigated the impact of this method in solving the classical 0-1 knapsack problem. According to the numerical results, we observe that the cascading-tree approach increases the performance of the algorithm. This approach has already been used in solving the classical \textit{Traveling Salesman Problem} (TSP) \citep{Book.HMM.1973}.
However, as future research directions, the cascading-tree branching can be combined by existing most efficient branch-and-bound algorithms for solving the large-scale instances of the knapsack problem or, possibly, other combinatorial optimization problems, which are out of the scope of the current memorial paper.
%


\nocite{*}

\bibliographystyle{itor}
\bibliography{bibliography}

\end{document}